# AUTOMATED INTEGRATION OF INFRASTRUCTURE COMPONENT STATUS FOR REAL-TIME RESTORATION PROGRESS CONTROL: CASE STUDY OF HIGHWAY SYSTEM IN HURRICANE HARVEY

Yitong Li
Wenying Ji

Fengxiu Zhang

Department of Civil, Environmental, and
Infrastructure Engineering
George Mason University
Fairfax, VA 22030, USA

Schar School of Policy and Government

George Mason University
Arlington, VA 22201, USA

## ABSTRACT

Following extreme events, efficient restoration of infrastructure systems is critical to sustaining community lifelines. During the process, effective monitoring and control of the infrastructure restoration progress is critical. This research proposes a systematic approach that automatically integrates component-level restoration status to achieve real-time forecasting of overall infrastructure restoration progress. In this research, the approach is mainly designed for transportation infrastructure restoration following Hurricane Harvey. In detail, the component-level restoration status is linked to the restoration progress forecasting through network modeling and earned value method. Once the new component restoration status is collected, the information is automatically integrated to update the overall restoration progress forecasting. Academically, an approach is proposed to automatically transform the component-level restoration information to overall restoration progress. In practice, the approach expects to ease the communication and coordination efforts between emergency managers, thereby facilitating timely identification and resolution of issues for rapid infrastructure restoration.

## 1    INTRODUCTION

Infrastructure systems (e.g., transportation, energy, water, and telecommunication) are vulnerable to extreme events (e.g., hurricanes, wildfires, and severe storms) (Homeland Security 2013). Following extreme events, the rapidly changing environment continuously causes infrastructure damage (e.g., highway inundation (Chen et al. 2020), public service delay or discontinuity (Miao et al 2018; Zhang 2021), power outage (Chen et al. 2021), and water pipeline failure (Yu et al. 2021)), impacting community lifelines and hampering emergency response (Homeland Security 2013, 2019). Given the transboundary nature of extreme events and their impacts, infrastructure service continuity depends not only on the ability of infrastructure systems to remain functional (Bruneau et al. 2003), but also on stakeholders' ability to effectively manage the restoration process (Gomez et al. 2019; Li and Ji 2021; Li et al. 2022). Even though numerous tools and methodologies have been proposed to analyze the performance of the physical system by identifying damaged infrastructure components and measuring changes of system functionality (Hernandez-Fajardo and Dueñas-Osorio 2013; Yu et al. 2021), research has rarely focused on supporting the monitoring and control of restoration progress.

The monitoring and control of infrastructure restoration progress is critical as it enables emergency managers to identify issues and take proactive actions in a timely manner (National Academies of Sciences, Engineering, and Medicine 2017). Following an extreme event, emergency managers acting from dispersed locations at various levels of governments make plans about restoration task prioritization and resource allocation (National Academies of Sciences, Engineering, and Medicine 2017; Gomez et al. 2019). Along with the changes of component restoration status, the plans need to be adjusted accordingly. In practice, due to the uncertain characteristics of extreme events (e.g., event intensity, damage severity, and resource availability) (Yu and Baroud 2020; Yu et al. 2021; Zhang and Maroulis, 2021) and challenges to interorganizational coordination (Comfort 2002; Zhang and Welch, 2021), infrastructure restoration progress is subject to complex dynamic changes. Changes in event impacts, demands and restoration progress require real-time information and resource exchange (Li and Ji 2020; Cao et. al. 2022), and consequently burden the communication and coordination efforts among responding agencies (National Academies of Sciences, Engineering, and Medicine 2017). The distributed information and resources as well as parallel operations in response to an extreme event calls for an approach that integrates component-level restoration status to support the monitoring and control of infrastructure restoration progress. To achieve this, our proposed approach addresses two research questions. First, how to transform the component-level restoration status to system-level infrastructure restoration progress? Second, how to synchronize restoration progress forecasting with updated restoration status at the component level?

The objective of this research is to propose a systematic approach that automatically integrates component-level restoration status to achieve real-time forecasting of infrastructure restoration progress. While the approach is mainly proposed for transportation infrastructure restoration following flooding, the concept can easily be generalized to different types of infrastructure systems with necessary modifications. The proposed approach is composed of the following steps. First, the topology and restoration status of highway infrastructure are represented using networks. Then, efficiency, a topological-based network metric, is used to measure highway functionality. Based on the highway functionality measured at different time steps, beta cumulative distribution function (CDF) is used to model infrastructure restoration progress. Lastly, as the new restoration status is identified, the information is used to update restoration progress forecasting using Bayesian inference with Markov Chain Monte Carlo (MCMC) and earned schedule method. Academically, the proposed approach is capable of automatically integrating component-level restoration status to achieve up-to-date restoration progress forecasting. In practice, this approach can help emergency managers quickly take proactive actions to avoid issues that may delay infrastructure restoration.

The remainder of this paper is organized as follows. In the next section, existing modeling approaches on infrastructure restoration are introduced. Then, examinations of the feasibility of using earned value analysis in linking the component-level restoration status and system-level restoration progress forecasting are provided. After that, a systematic approach that automatically integrates component-level restoration status to obtain up-to-date restoration progress forecasting is introduced step by step. In the end, contributions and future work are concluded.

## 2    LITERATURE REVIEW

Extreme events typically span organizational, geographical and jurisdictional boundaries and require swift concerted infrastructure restoration under deep uncertainty and high complexity (Zhang et al., 2018; Ansell et al., 2010). The widespread geographical distribution of transportation infrastructure and complex infrastructure interdependence significantly complicates the communication and coordination among a myriad of responding agencies. In conducting distributed and parallel operations, most agencies would be restoring a certain segment or component of the infrastructure, while lacking a full real-time view of the entire system. The restoration process becomes more complicated when accounting for the fragmented ownership, operation and maintenance of transportation infrastructure and the diversity of routines and operating practices among the involved actors. The distributed and somewhat fragmented response system requires a reliable approach to synthesizing information and monitoring and controlling the infrastructure restoration progress.

Aiming at facilitating efficient infrastructure restoration, numerous approaches, such as empirical (Utne et al. 2011; Kjølle et al. 2012), agent-based (Pumpuni-Lenss et al. 2017; Rasoulkhani et al. 2020), and network-based approaches (Hernandez-Fajardo and Dueñas-Osorio 2013; Yu et al. 2021) have been proposed. Among these approaches, the network-based approach has been deemed the most suitable because it can model complex interactions among various infrastructure components (Hernandez-Fajardo and Dueñas-Osorio 2013). This is achieved by presenting nodes as infrastructure components and edges as interdependent components. Despite being useful, existing approaches have primarily focused on physical dimensions of infrastructure systems and neglected the difficulties for emergency management agencies to interpret the model results. Quite often, model results need to be further interpreted and analyzed for the agencies to quickly grasp the overall restoration progress and exercise effective restoration progress control.

In project control, earned value is a method that is widely used for the management of project cost and schedule forecasting (Vandevoorde and Vanhoucke 2006). Essentially, this method is concerned with three values: the planned, earned, and actual values. These values are obtainable from S-curve, which shows changes of cumulative project progress (e.g., cost) over time (Project Management Institute 2000). S-curve assumes that tasks that have been finished or will be finished by a certain time step during the project execution phase resembles an S-shaped curve. From S-curve, the value at a certain time step can be identified and further used to estimate the value at completion. This is achieved by summing the actual value completed to date and the estimated value for the remaining work. In the context of infrastructure restoration, earned value method provides a feasible way to link the detailed component-level restoration status to system-level restoration progress forecasting. In detail, each of the damaged infrastructure components can be viewed as a task, and restoration of each task continuously contributes to system functionality. Overall, infrastructure restoration progress can be modeled as an S-curve that illustrates changes of system functionality over time. By modeling infrastructure systems as networks and applying global network metrics, the component-level restoration status can be integrated to evaluate system functionality. Once changes of system functionality at various time steps are obtained, earned value method can be applied to forecast future restoration progress.

## 3   METHODOLOGY

Hurricane Harvey is a category 4 hurricane that caused widespread damage to Southeast Texas. The Hurricane resulted in tremendous rainfall, leading to severe flooding that damaged the state's transportation system (Blake and Zelinsky 2018). As a result, emergency managers' ability to perform response actions (e.g., accessing disaster sites, allocating emergency relief personnel and supply, and rescuing victims) were impeded. Among various impacted counties, Harris County is the most damaged and therefore is investigated in this research. To identify the incident location and highway restoration status, datasets that describe highway geography (Texas Department of Transportation 2022) and highway high-water incidents (Chen et al. 2020) were used. In detail, the highway geography specifies the location of roadways in Harris County. It describes detailed information of highway roads and road intersections. The highway high-water incidents contain 328 incident reports that range from August 25[th], 2017 23:41:00 to September 5[th], 2017 13:15:00. This information was collected by Texas Department of Transportation. Here, each of reports documents the coordinate (longitude and latitude) of where the incident was observed, the start time when the incident was observed, and the end time when the incident was cleared. By mapping the incident coordinates to the highway geography, an overview of the highway system (indicated by pink lines) and the spatial distribution of daily incidents (indicated by red dots) are shown in Figure 1. Here, highwater incidents gradually appeared on August 25[th], 2017 and August 26[th], 2017. Then, a significant number of incidents was observed from the days of August 27[th], 2017 to August 30[th], 2017. Following this period, the incidents gradually disappeared as a result of restoration efforts.

The process of integrating component restoration status to obtain up-to-date restoration progress forecasting is illustrated in Figure 2. In detail, changes of highway restoration status are represented as network models with various network topologies. Once modeled, network efficiency is used to compute highway functionality. Using changes in highway functionality over time, infrastructure restoration progress is modeled as a beta CDF. Lastly, as new restoration status becomes available, the corresponding network functionality is automatically computed and used to obtain up-to-date restoration progress forecasting, which is achieved using Bayesian inference with MCMC and earned schedule method. The detailed explanations of each of the steps are given in the following sections.

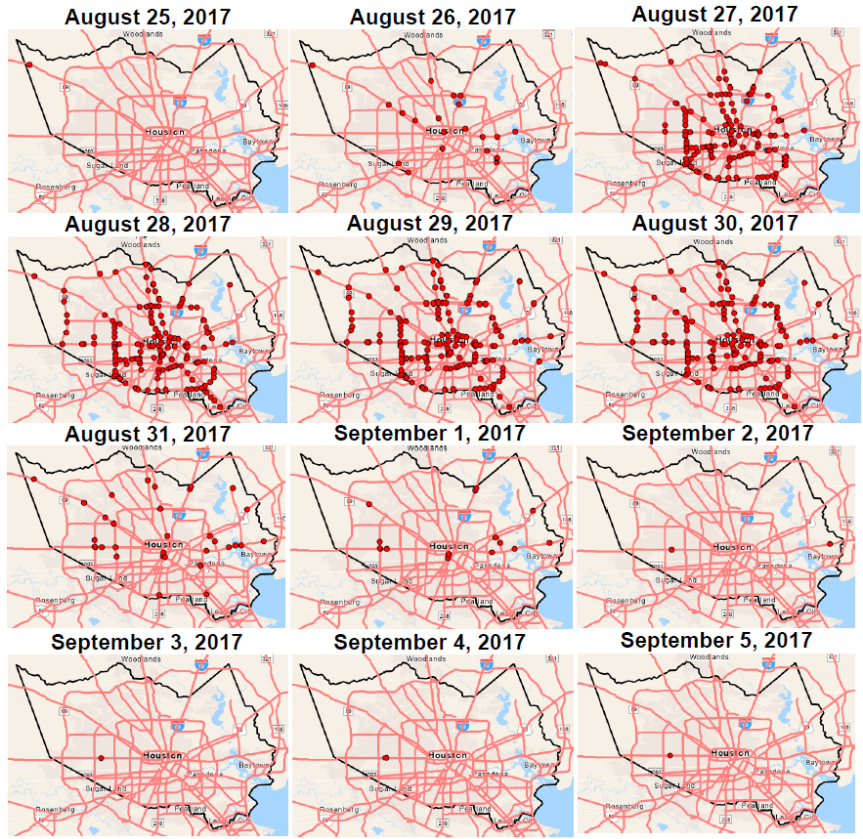

Figure 1: Daily highway incidents.

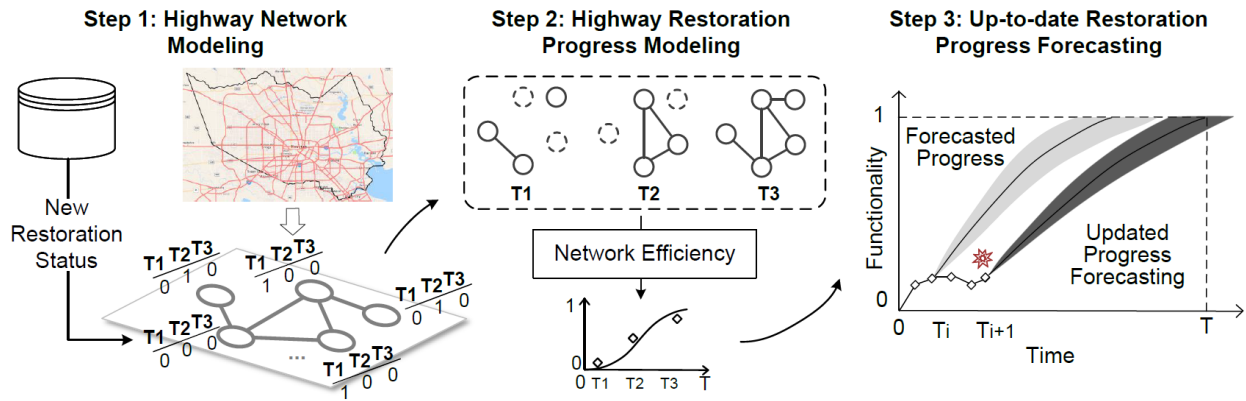

Figure 2: Research framework.

## 3.1    Highway Network Modeling

The highway infrastructure system is modeled as a network in which nodes representing highway intersections and edges representing highways (illustrated in Figure 3). ArcGIS, which is a geographical information system software, is used to visualize the geography of the highway system (ArcGIS 2022).

Based on the visualization, the intersections are manually labeled. In summary, the processed highway network has 272 nodes and 390 edges.

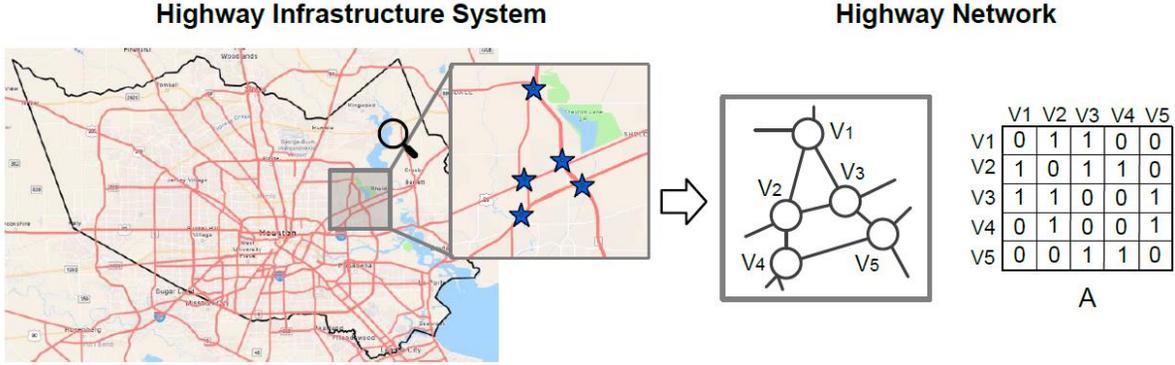

Figure 3: Extraction of highway network topology.

Following a disruptive event, some of the highway intersections (i.e., nodes) and highways (i.e., edges) may be damaged, which can impact the functionality of the other interdependent highways. In this research, the damaged network is modeled as follows. If the incidents happened on edges, the damaged edges will be removed. If the incidents happened on nodes, all edges connected to the damaged nodes will be removed. In the end, the generated network is a subnetwork of the highway network prior to the disruption. To represent the dynamic changes of highway restoration status, the highway network is mathematically defined as follows. Let $V^{(t)} = \{v_1, v_2, \ldots, v_n\}$ denotes a set of n intersections (i.e., nodes) in a highway network $G^{(t)}$ at time step $t$. The adjacency matrix $A^{(t)} = \left(a_{ij}\right)^{n \times n}$ is used to denote the topology of the highway network, where $a_{ij} = a_{ji} = 1$ when two intersections are connected, and $a_{ij} = a_{ji} = 0$ when they are not. Within the highway network, each component (i.e., highway intersections and highways) is affiliated with the binary number $N \in \{0,1\}$ to denote network damage conditions at time step $t$, in which 0 indicates functional components and 1 indicates damaged components (demonstrated in Figure 2 Step 1).

## 3.2  Highway Restoration Progress Modeling

Once damages on a highway network at different time steps are obtained, they are used to compute highway functionality. Network efficiency is selected to measure highway functionality. Mathematically, network efficiency measures the shortest path length between paired nodes in a network (Latora and Marchiori 2001). It is proposed based on the assumption that the more distant a pair of nodes are in a network, the less efficient the network will be. In the context of highway systems, network efficiency measures how fast passengers can move from one location to another location. The efficiency for network $G$ with $N$ nodes is

$$E(G) = \frac{1}{N(N-1)} \sum_{i \neq j \in G} \frac{1}{d_{ij}}, \tag{1}$$

where $N$ is the total number of interactions in the network and $d_{ij}$ is the shortest length (i.e., path) between intersections $i$ and $j$.

Infrastructure restoration progress can be described by changes of infrastructure functionality over time (Zorn and Shamseldin 2015). To model infrastructure restoration progress, statistical-curve fitting approaches are widely used due to its capability of generalization. At the core of the approach is to identify a mathematical function that well represents infrastructure restoration progress. In this research, beta CDF $Beta(a, b, L, U)$ is selected because (1) it has closed-form boundaries to represent the start and finish of infrastructure restoration progress; (2) it is a monotonic function that illustrates the continuous increase of infrastructure functionality as a result of restoration efforts; and (3) it has flexible shapes to model changes of restoration progress. In this research, since the start time of infrastructure restoration is always 0, beta CDF is simplified to three parameters $Beta(a, b, 0, U)$.

### 3.3 Up-to-date Restoration Progress Forecasting

As highway restoration progresses, new restoration status will become available for deriving the latest changes of highway functionality. Once obtained, these data are used to update restoration progress forecasting. This is achieved by first updating the restoration progress and then using its output to forecast future restoration progress. Bayesian inference is selected to perform the updating because it provides a systematic way of updating information when more information becomes available (Gelman et al. 2013). Since restoration progress is modeled as Beta CDF ($Beta(a, b, 0, U)$), the updating essentially narrows down to the updating of three parameters $a$, $b$, and $U$. The general process of updating beta CDF using Bayesian inference has been discussed in the Bayesian adaptive forecasting model proposed by Kim and Reinschmidt (2009). In short, the model uses Monte Carlo integration to numerically approximate marginal posterior distributions of parameters $a$, $b$, and $U$. However, Monte Carlo simulation is not guaranteed to generate sufficient samples to represent the posterior distribution (Qian et al. 2003). As an improvement, Markov chain Monte Carlo (MCMC), which has been demonstrated as a more efficient and reliable method, is used (Chib and Greenberg 1995; Hoff 2009). Doing so allows us to obtain the parameters $a$, $b$ and $U$ to indicate the updated restoration progress.

Once the restoration progress is updated, it is used to forecast future restoration progress (shown as Figure 4).

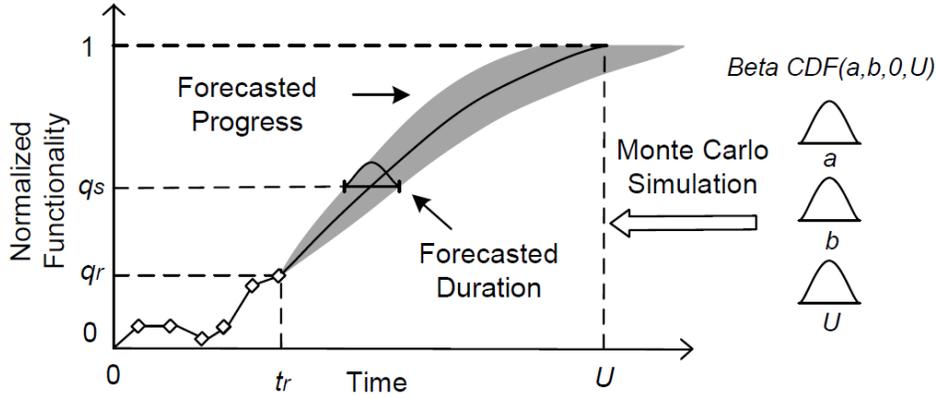

Figure 4: Up-to-date restoration progress forecasting.

At the core of the progress forecasting is to estimate the duration needed to restore the infrastructure functionality to a specified level, in which the specified level ($q_s$) needs to be higher than the restored functionality ($q_r$) but lower than the original functionality (1 after normalization). In earned value analysis, earned schedule method (Vandevoorde and Vanhoucke 2006) has been demonstrated to be the most reliable method for duration forecasting. The basic idea of the method is to estimate the duration at completion based on the current schedule performance index. Specializing the earned schedule method, the forecasted restoration duration required to restore the infrastructure system to $q_s$ is

$$Forecasted\ Duration = t_r + \frac{BetaCDF(q_s) - BetaCDF(q_r)}{\frac{BetaCDF(q_r)}{t_r}}, q_r \leq q_s \leq 1. \quad (2)$$

In equation (2), $t_r$ represents the actual time taken to achieve the current system functionality, $BetaCDF(q_s)$ represents the planned time to achieve the specified functionality, and $BetaCDF(q_r)$ represents the planned time to achieve the current earned value. Since the planned functionality at completion is 100%, the earned value is simplified as $q_r$. Using equation (2), uncertainties associated with forecasted restoration duration are computed using Monte Carlo simulation (Raychaudhuri 2008). In detail, for each iteration, one sample is randomly selected from the distributions of the updated parameters $a$, $b$, and $U$. Then, earned schedule method (equation (2)) is applied to forecast the duration at completion. Repeating the iterations for $N$ number of times, a distribution of the forecasted duration can be obtained. Following the same process, the duration at completion is computed for all the specified functionality levels. The distributions corresponding to various functionality levels will form a region to illustrate the forecasted restoration progress.

## 4    RESULTS

### 4.1    Highway Functionality Measurement

The processed highway network in Harris County contains 272 nodes and 390 edges. To evaluate changes of highway functionality in a detailed granularity, the highway incident dataset was disaggregated to obtain incidents that happened on an hourly basis. Before processing, all data are rounded to the nearest hour using rules defined in the R package lubridate (Grolemund and Wickham 2011). Overall, the start time of the earliest incident (August 26th, 2017 00:00:00) and the end time of the latest incident (September 5th, 2017 13:00:00) were used to generate a time list, indicating the total number of hourly time steps. If the start and end times of an incident cover multiple time steps, then the incident is considered to be present in those time steps. In summary, 255 subnetworks, representing the damaged highway networks in different hours, were generated. Using the damaged networks, equation (1) was used to compute changes of highway efficiency over time. The results are illustrated in Figure 5.

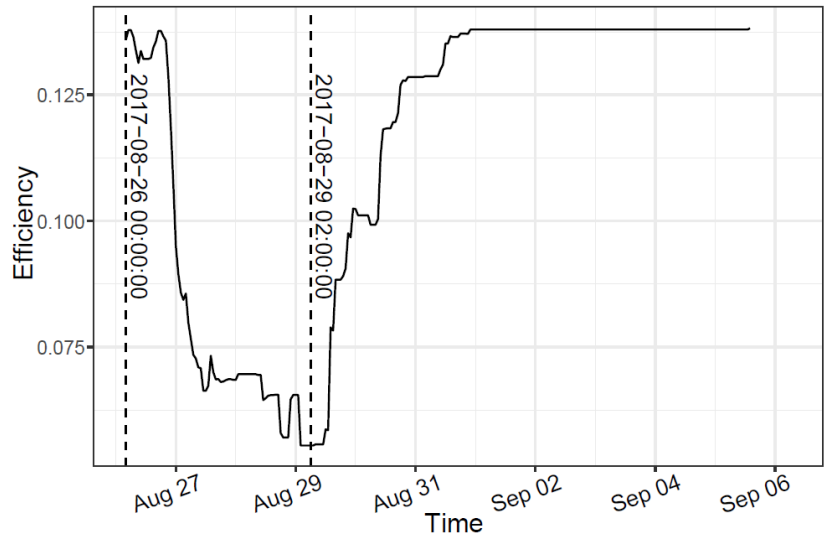

Figure 5: Changes of network efficiency over time.

Prior to the disruption, the highway efficiency is 0.138. Following the disruption, the efficiency decreases and hits the lowest value (0.0555) at time 2017-08-29 02:00:00. Following this time step, the restoration efforts started, and the highway efficiency was restored to the original value.

## 4.2    Highway Restoration Progress

Restoration progress is identified as changes of highway functionality within the time interval starting from the time step with the lowest functionality to the time step when the restoration is completed. In practice, the exact restoration completion time is difficult to identify. To address this issue, one commonly used approach is to specify a functionality level and use the corresponding time as the restoration completion time (Zorn and Shamseldin 2015). Here, 99% was selected because no significant changes in functionality were observed following this time step. In summary, the restoration progress ranges from 2017-08-29 02:00:00 to 2017-08-31 21:00:00, and the total highway restoration lasted 68 hours. Once the start and end time steps were identified, the functionality values in between were normalized from 0 to 1 to model the infrastructure restoration progress as Beta CDF. In summary, the restoration progress took 68 hours and the processed infrastructure restoration progress is shown in Figure 6, in which each of the dots indicates the restoration progress completed by a certain hour.

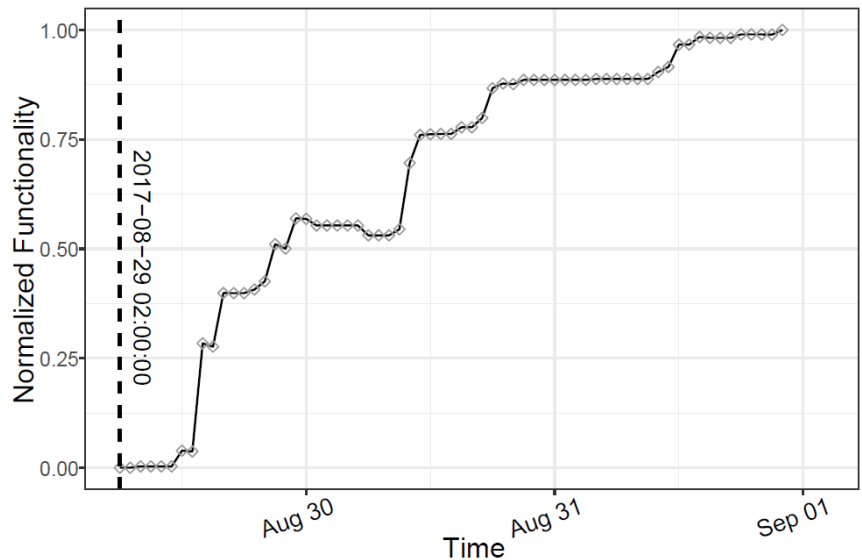

Figure 6: Highway restoration progress.

## 4.3    Up-to-date Restoration Progress Forecasting

To demonstrate the capability of the proposed approach in forecasting up-to-date restoration progress, the highway restoration progress was divided on a half-day basis (i.e., 12, 24, 36, 48, 60, and 68 hours). By the end of every 12 hours, the newly-identified restoration progress was used to update the planned restoration progress using MCMC. To perform MCMC, several inputs require specification. These are initial values of parameters $a$, $b$, and $U$, prior distributions of parameters $a$, $b$, and $U$ (i.e., planned restoration progress), proposal distributions of parameters $a$, $b$, and $U$, and the standard deviation of the likelihood distribution. The prior distributions can be determined from historical data and expert opinions. Here, historical data refer to infrastructure restoration progress (i.e., infrastructure functionality over time) collected from previous extreme events. When historical data is not available, priors can be specified by experts who have basic knowledge on a system and society's preparedness for event response (Cimellaro et al. 2010). The remaining inputs are associated with the accuracy and efficient of MCMC algorithm, which can be specified through systematic calibration (Hoff 2009). Once these inputs were specified, at each time step, a value was drawn from the proposal distribution and the new value was either accepted or rejected based on an acceptance ratio. In the end, representative samples that illustrate posterior distributions of $a$, $b$, and $U$ were obtained, indicating the updated restoration progress. Based on the results, earned schedule method (equation (2)) and Monte Carlo simulation were applied to forecast future restoration progress. A summary of the specified inputs is provided in Table 1. Notably, the proposal distributions are defined as normal distributions with the mean specified as the current value in the chain and the standard deviation (S.D.) specified as the values given in Table 1. The results of the forecasted restoration progress by the end of every 12 hours are shown in Figure 7.

Table 1: MCMC inputs.

| Input | Input Values |
|---|---|
| Initial value $(a, b, U)$ | $(1, 2, 70)$ |
| Prior distribution | $a \sim Normal(1, 0.100)$ $b \sim Normal(1.772, 0.177)$ $U \sim Normal(68, 6.8)$ |
| S.D. of the proposal distribution $(a, b, U)$ | $(0.0200, 0.0354, 1.36)$ |
| S.D. in the likelihood distribution | $3.911$ |

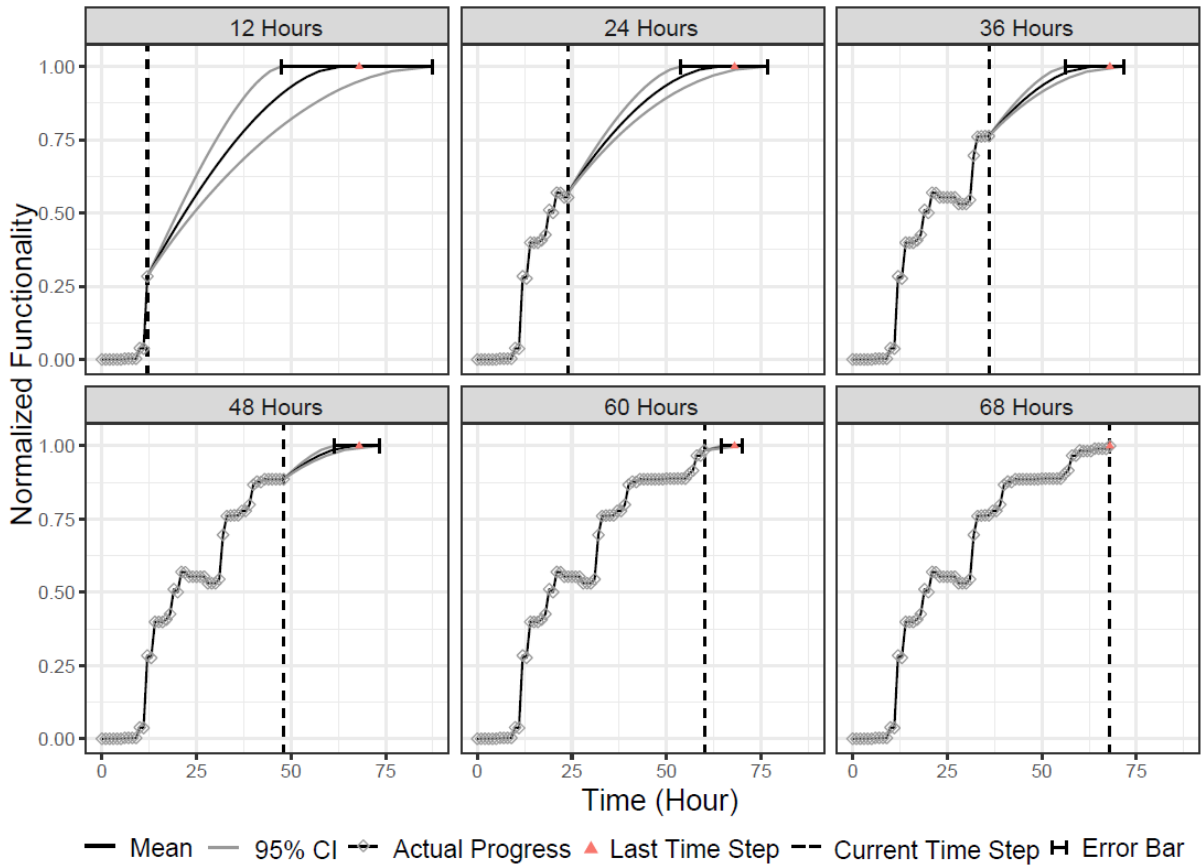

Figure 7: Up-to-date forecasting of highway restoration progress.

In each of the facets, the dashed line shows the current time step. Before the current time step, the connected line and gray dots show the actual restoration progress. After the current time step, the gray and black lines show the 95% credible interval and the mean of the forecasted restoration progress, respectively. The error bar on the upper right corner represents the 95% credible interval (CI) of the forecasted restoration duration at completion. Lastly, the red triangle represents the actual restoration duration at completion. For all the forecasted results, the actual restoration duration falls within the bounds of the 95% CI, which validates the accuracy of the forecasted results.

In reality, the results can be interpreted as follows. By the end of every 12 hours, the component restoration status is collected and automatically integrated to obtain the up-to-date forecasting of infrastructure restoration progress. Immediately, emergency managers are able to oversee the future restoration progress and analyze the restoration progress to adjust restoration plans. In addition, the up-to-date progress forecasting also ensures that a common understanding of infrastructure restoration progress is shared among emergency managers and high-level management agencies. During highway infrastructure restoration at Harris County, a significant functionality increase was observed in the first 24 hours. In the next 24 hours, the restoration progress slowed down. As a result, the restoration curve slightly shifts toward the right, indicating that the progress for future restoration tasks is expected to decrease. At this time step, if proper proactive actions were taken, the restoration progress may be expedited.

## 5    CONCLUSION

Efficient infrastructure restoration is essential for stabilizing community lifelines. A large body of research has been proposed to facilitate decision-making focused on physical dimensions of infrastructure systems. While understanding the performance of the physical system is essential, the monitoring and control of infrastructure restoration progress is also critical. To that end, this research proposes a systematic approach that automatically integrates component-level restoration status to achieve real-time forecasting of the system-level restoration progress. While the approach is designed for highway infrastructure restoration following Hurricane-induced flooding, the concept can be applied to different types of infrastructure systems under various disaster contexts. The systematic approach first models highway systems as networks and use network efficiency to indicate highway functionality. Once values of highway functionality at various time steps are computed, Bayesian inference with MCMC and earned schedule method are used to forecast up-to-date restoration progress.

In practice, the proposed approach is expected to ease the communication and coordination efforts among emergency managers. Specifically, once new restoration status is identified, the overall restoration progress forecasting is automatically derived which can support emergency managers to perform timely identification and resolution of issues. Currently, the modeled highway network solely considers highway topology. In practice, however, highway functionality is influenced by other factors, such as volumes and directions of traffic flows. In addition, although the proposed approach is able to support emergency managers in quickly identifying potential issues and adjusting restoration plans (e.g., task prioritization and resource allocation), detailed relations between the identified issues and restoration plans are unclear. Future efforts will need to link restoration tasks with restoration progress forecasting to better support emergency managers in making actionable plan adjustments in a timely manner.

## AUTHOR BIOGRAPHIES

**YITONG LI** is a Ph.D. candidate in the Department of Civil, Environmental & Infrastructure Engineering, George Mason University. Yitong's current research area focuses on dynamic modeling of infrastructure restoration progress and construction simulation input modeling. Her e-mail address is yli63@gmu.edu.

**FENGXIU ZHANG** is an assistant professor in Schar School of Policy and Government, George Mason University. Dr. Zhang received her Ph.D. in Public Administration and Policy from Arizona State University. Dr. Zhang's research interests include decision making under risk and uncertainty, extreme events, climate change adaptation, organizational adaptation and resilience, critical infrastructure protection and technology in government. Her email address is fzhang22@gmu.edu.

**WENYING JI** is an assistant professor in the Department of Civil, Environmental & Infrastructure Engineering, George Mason University. Dr. Ji received his PhD in Construction Engineering and Management from the University of Alberta. Dr. Ji is an interdisciplinary scholar focused on the integration of advanced data analytics and complex system modeling to enhance the overall performance of infrastructure systems. His e-mail address is wji2@gmu.edu.